\documentclass[conference]{IEEEtran}
\IEEEoverridecommandlockouts

\usepackage{cite}
\usepackage{amsmath,amssymb,amsfonts}
\usepackage{algorithmic}
\usepackage{graphicx}
\usepackage{textcomp}
\usepackage{xcolor}
\usepackage{subcaption}
\usepackage{tabularx}
\usepackage{pifont}
\usepackage{multirow}
\def\BibTeX{{\rm B\kern-.05em{\sc i\kern-.025em b}\kern-.08em
    T\kern-.1667em\lower.7ex\hbox{E}\kern-.125emX}}
\begin{document}

\title{Understanding the Representation of Older Adults in Motion Capture Locomotion Datasets
\thanks{This work is supported partly by NSERC sMAP CREATE and NSERC Discovery grants.}
}
\author{\IEEEauthorblockN{1\textsuperscript{st} Yunkai Yu}
\IEEEauthorblockA{\textit{Dept. of Computing and Software} \\
\textit{McMaster University}\\
Hamilton, Canada \\
yu284@mcmaster.ca}
\and
\IEEEauthorblockN{2\textsuperscript{nd} Yingying Wang}
\IEEEauthorblockA{\textit{Dept. of Computing and Software} \\
\textit{McMaster University}\\
Hamilton, Canada \\
wang840@mcmaster.ca}
\and
\IEEEauthorblockN{3\textsuperscript{rd} Rong Zheng}
\IEEEauthorblockA{\textit{Dept. of Computing and Software} \\
\textit{McMaster University}\\
Hamilton, Canada \\
rzheng@mcmaster.ca}
}

\maketitle

\begin{abstract}
The Internet of Things (IoT) sensors have been widely employed to capture human locomotions to enable applications such as activity recognition, human pose estimation, and fall detection. Motion capture (MoCap) systems are frequently used to generate ground truth annotations for human poses when training models with data from wearable or ambient sensors, and have been shown to be effective to synthesize data in these modalities. However, the representation of older adults, an increasingly important demographic in healthcare, in existing MoCap locomotion datasets has not been thoroughly examined. This work surveyed 41 publicly available datasets, identifying eight that include older adult motions and four that contain motions performed by younger actors annotated as old style. Older adults represent a small portion of participants overall, and few datasets provide full-body motion data for this group. To assess the fidelity of old-style walking motions, quantitative metrics are introduced, defining high fidelity as the ability to capture age-related differences relative to normative walking. Using gait parameters that are age-sensitive, robust to noise, and resilient to data scarcity, we found that old-style walking motions often exhibit overly controlled patterns and fail to faithfully characterize aging. These findings highlight the need for improved representation of older adults in motion datasets and establish a method to quantitatively evaluate the quality of old-style walking motions.
\end{abstract}


\section{Introduction}
The Internet of Things (IoT) has made locomotion data more accessible, facilitating innovations in applications ranging from human activity recognition to fall detection and exoskeleton design. A motion capture (MoCap) system is the gold standard for recording locomotion data with high precision in controlled lab environments. MoCap data has been widely used to annotate data from IoT sensors such as mmWave radars and inertial measurement units in human motion analysis \cite{zhang2024super}, or synthesize data from other modalities \cite{hao2022cromosim}. Data-driven models are increasingly important in motion analysis, often trained directly on MoCap datasets without relying on hand-crafted features. As such, the characteristics of these datasets, such as the variety of motions and the demographic composition, directly influence model performance. 

Considering the growing older population worldwide, it is important to understand the coverage and quality of old adult motions in public datasets. David \emph{et al.} \cite{locomotionreview} surveyed 93 locomotion datasets and reported number of subjects, motor skills, application domains, sensory systems, and data formats. They found that existing datasets lack detailed demographic information of the participants. Few studies have systematically examined how well existing MoCap datasets capture the motion patterns of older adults.

This work aims to answer two research questions. \textit{1) What is the availability of older adult locomotion data in existing public MoCap datasets?} and \textit{2) Do existing old-style locomotion data performed by actors have sufficient fidelity to accurately reflect age-related differences?} To answer the first question, we surveyed 41 datasets collected by clinical and engineering communities and analyzed their demographic coverage, motor skills, motion varieties, and whether the data includes full-body or partial-body motions. Among these, only eight datasets are known to include older adults. In total, 121 of the 385 participants are older adults, but full-body motion data is available for 75 of them. The motor skills in full-body motions are limited to walking and standing for older adults.

To answer the second question, we further analyzed the old-style forward walking motions in four of the 41 datasets. We selected gait parameters known to capture age-related differences and used them as quantifiable metrics to evaluate the fidelity of old-style motions. Our analysis reveals that only around 12 minutes of old-style forward walking are available, and the recorded movements often fail to accurately represent age-related gait patterns.

In summary, the main contributions of this work include:

\begin{itemize}
    \item It conducted the first survey of public MoCap locomotion datasets focusing on demographic coverage, motor skills, motion varieties, and the anatomical completeness of MoCap for older adults.
    \item It proposes key objective metrics for assessing the fidelity of old-style motions with respect to their ability to reflect age-related differences. 
    \item It systematically reveals for the first time the low availability of high-quality locomotion data for older adults in public datasets. 
\end{itemize} 

The rest of the paper is organized as follows. Section~\ref{sec:pre} introduces key terminologies used in subsequent discussions. Section~\ref{sec:method} presents a qualitative analysis of the datasets discussed in this paper. Section~\ref{sec:fidelity} presents our quantitative assessment method and assesses the fidelity of old-style motions in existing MoCap locomotion datasets. Section~\ref{sec:conclusion} concludes the paper with key findings and future directions.

\section{Preliminaries}\label{sec:pre}
\textit{Clinical datasets} in this work are interested in locomotion profiles for specific groups. In these datasets, motion data is collected from participants of interest in controlled lab environments. In addition to MoCap data, some datasets, e.g. \cite{tamayavan2023full}, include ground reaction force and electromyography (EMG) data to understand human dynamics. 

\textit{General-purpose datasets} are mainly collected to support applications in video games, animation, simulation, and other fields that rely on human motions. These datasets aim to capture a wide range of motor skills, enabling researchers and practitioners to create realistic and adaptable animations. Typically, the motions in these datasets came from campus volunteers or professional actors, who are directed to perform various motor skills with precision and consistency.

\textit{Motor skills} refer to the ability to perform specific functional movements that involve coordination of muscles. 

\textit{Style} is the manner in which a motor skill is performed. It conveys emotions, personality, and intentions. For example, Xia \emph{et al.}'s dataset \cite{xia2015dataset} includes eight distinct styles: angry, childlike, depressed, neutral, old, proud, sexy, and strutting. In contrast, Ian \emph{et al.} \cite{100style2022} introduced the 100style dataset, which contains 100 performative styles, with \textit{old} being one of them. The BFA \cite{aberman2020unpaired} dataset also guided actors to perform motions to resemble older adults. Motions labeled with such style annotations are referred to as stylized motions. Old style is subjective and can be performed using different movement patterns depending on the performers' interpretation.

\textit{Motion variety} refers to the range and diversity of movement patterns. In this work, we focus on \textit{natural and age-related movement patterns} within the same motor skills, excluding artificial variations sourced from unnaturalness or age-irrelevant factors. Ambiguity in the guideline and procedure designed to collect motions results in variations even within the same motor skill, which are excluded from consideration. For example, the motion description of the $0005{\_}{Walking001}$ in the SFU MoCap dataset \cite{SFUMocap} is ``walking forwards''. However, the walking motion clip includes instances of turning around. Similarly, a clip labeled as \textit{walking in straight lines with increasing stride length} includes an intentional modification of gait dynamics, which constitutes an artificial variation rather than a natural difference \cite{SFUMocap}. Moreover, participants may not have rigorously followed the same experimental setup when performing motions in different styles. For example, text annotations indicate that sequences $137\_{33}$ and $137\_{29}$ in the CMU MoCap dataset \cite{CMUMOCAP} correspond to the same motor skill. However, their walking trajectories differ significantly.

\section{MoCap Locomotion Datasets and Qualitative Analysis}\label{sec:method}
The MoCap locomotion datasets discussed in this work were identified through a systematic literature search conducted through the Web of Science, focusing on research published between 2021 and 2025. The search was done using the query: TS=(``human body motion'' OR ``locomotion'' OR ``human activity'') And TS=(``physics'' OR ``motion capture'') AND TS=(``datasets''). Among selected studies, public MoCap datasets containing common locomotion captured by optical markers are included. If a MoCap dataset is constructed from multiple sources, only the primary source datasets that contain locomotion data will be included. We obtained 19 clinical and 22 general-purpose datasets, comprising over 995 participants, including 121 identified as older adults, and four containing old-style motions. 

\subsection{Clinical Datasets}
\begin{table*}
    \centering
    \begin{tabularx}{\textwidth}{cccXc}
        \hline
        Clinical Datasets&Older Adults/Total Participants&Body Parts&Motor Skills\\
        \hline    
        Falisse \emph{et al.} \cite{Falisse}&0/8&Full body&gait, squat, stair descent/ascent, sit-to-stand-to-sit, and squat jump.\\
        Wang \emph{et al.} \cite{wang2023wearable}&0/9&Lower limbs and trunk& walking, running, vertical jump, squat, lunge, and single-leg landing.\\
        
        Van der Zee \emph{et al.} \cite{vanderzee2022biomechanics}&0/10&Lower limbs&walking across different speeds, step lengths/frequencies/widths, and walking conditions.\\
        Reznick \emph{et al.} \cite{Reznick}&0/10&Lower limbs&walking at multiple inclines and speeds, running at multiple speeds, walking and running with constant acceleration, and stair ascent/descent with multiple stair inclines.\\
        Hamner \emph{et al.} \cite{HAMNER2013780}&0/10&Full body& running at different speeds.\\
        Uhlrich \emph{et al.} \cite{uhlrich2023opencap}&0/10&Full body&four activities (squats, sit-to-stand, drop vertical jump, and walking) with varied kinematic patterns.\\
        Pequera \emph{et al.} \cite{pequera20204134978}&0/14&Full body&walking, running and unilateral skipping on a treadmill.\\
        Tan \emph{et al.} \cite{tan2022strike}&0/16&Feet&running with three strike patterns in four conditions (footware types and two running speeds).\\
        Carmargo \emph{et al.} \cite{CAMARGO2021110320}&0/22&Lower limbs&level-ground/treadmill walking, stair ascent/descent, and ramp ascent/descent.&\\
        Carter \emph{et al.} \cite{carters}&0/50&Full body&running at different speeds and gradients.\\
        Horst \emph{et al.} \cite{horst2019explaining}& 0/57 &Full body&barefoot walking at a self-selected speed on a 10-m path.\\
        Li \emph{et al.} \cite{Li2021}&1/2&Lower limbs&walking.\\
        Lencioni \emph{et al.} \cite{Lencioni2019HumanKK}&3/50&Full body&level-ground walking, treadmill walking, ramps, and stairs.\\
        Moore \emph{et al.} \cite{moore2015}&3/50&Full body&normal walking and perturbed walking.\\ 
        Toronto Older Adults Gait Archive \cite{mehdizadeh2022toronto}& 14/14 &Lower limbs&walking in 60 seconds.\\
        Fukuchi \emph{et al.} \cite{fukuchi2018public}&18/42&Lower limbs& overground and treadmill walking at a range of gait speeds\\
        Hafer \emph{et al.} \cite{HAFER2020109567KneeOA}& 20/30&Lower limbs&overground walking at three self-selected speeds.\\
        Santos \emph{et al.} \cite{santos2017data}&22/49&Full body&standing still.\\ Tamaya \emph{et al.} \cite{tamayavan2023full}&40/138&Full body&walking at preferred speeds.\\
        \hline
    \end{tabularx}
    \caption{Basic information of clinical locomotion datasets.}
    \label{tab:clinic}
\end{table*}
\subsubsection{Dataset Size and Demographics}
Most of the clinical MoCap locomotion datasets recruited no less than 10 participants, as shown in TABLE~\ref{tab:clinic}. The participation of older adults is limited. For example, no older adults were recruited to perform locomotions \cite{Falisse,wang2023wearable,vanderzee2022biomechanics,Reznick,HAMNER2013780,uhlrich2023opencap,pequera20204134978,CAMARGO2021110320,carters,horst2019explaining}. 
Additionally, the dataset size for full-body locomotion data is limited. Clinical datasets often show uneven representation of different body parts, with the majority of full-body motion data coming from younger adults. Some datasets feature full body motions \cite{Falisse,HAMNER2013780,uhlrich2023opencap,pequera20204134978,carters,horst2019explaining,Lencioni2019HumanKK,moore2015,santos2017data,tamayavan2023full}, whereas some \cite{wang2023wearable,vanderzee2022biomechanics,Reznick,tan2022strike,CAMARGO2021110320,Li2021,mehdizadeh2022toronto,fukuchi2018public,HAFER2020109567KneeOA} focus solely on lower limb movements. This may be because the datasets were designed to assess the mobility of a specific population, focusing primarily on mobility metrics involving trunk and lower limbs, such as knee angles and steps. As a result, upper limb markers were omitted, and arm movements, despite their contribution to balance control \cite{mehdizadeh2022toronto,wang2023wearable}, were not recorded. 

\subsubsection{Motion Variety Within the Same Motor Skills}
In datasets where older adults account for a reasonable proportion of participants \cite{Li2021,mehdizadeh2022toronto,fukuchi2018public,HAFER2020109567KneeOA,santos2017data,tamayavan2023full}, the recorded motor skills are typically limited to walking and standing still. The absence of other types of locomotion may indicate the frailty nature of the group, as motions such as squat jumps and fast running can be too challenging for older participants. 

Within the category of walking, data collection protocols often require participants to perform the task repeatedly under standardized conditions. This consistency ensures that variations among motion clips primarily reflect natural inter-individual differences. For example, in \cite{mehdizadeh2022toronto}, all the recorded motions involve 60 seconds of walking in nursing homes, with participants keeping their arms relaxed and swinging naturally. The motion variety in the dataset can be quantified by the total minutes of walking recorded and the number of participants. 

\subsection{General-Purpose Datasets}
\subsubsection{Dataset Size and Demographics}
For the majority of the general-purpose datasets listed in TABLE~\ref{tab:nAllCV}, only a small portion of participants are known to be older adults. Age information is not released in datasets such as TotalCapture \cite{TotalCaptureTrumbleBMVC2017}, HDM05 \cite{BMLRUB}, CMU MoCap \cite{CMUMOCAP}, and in datasets with motions in different styles \cite{xia2015dataset,aberman2020unpaired,100style2022}. Thus, the presence of older participants in these datasets is undecided. Only four of the datasets contain motions labeled as old-style. For the datasets that include age information, none include older participants. 

In addition to demographic imbalance, the quantity of old-style motion data is small. This is evident even in one of the most common forms of locomotion--walking forward. We counted the total forward walking minutes in old-style data and found that even the combined forward walking time in all datasets featuring old-style motions does not exceed 15 minutes, as shown in TABLE~\ref{tab:nOld}.
Moreover, other demographic factors relevant to locomotion performance have not been fully explored in these datasets. For example, aging is associated with an increased risk of chronic disease progression, such as knee osteoarthritis and Parkinson's disease that affect gaits \cite{HAFER2020109567KneeOA,mirelman2019gaitPArkison}. However, there is no evidence that the old style in the datasets represents individuals other than healthy older adults.

\begin{table}
    \centering
    \begin{tabular}{cccc}
    \hline
        Datasets&Participants&Older Adults&Old Style?\\
        \hline
        Transitions \cite{AMASS2019}&1&0&No\\ 
        MoCap-ULL \cite{estevez2015mocapull}&1&Unkown&Unknown\\
         The Edinburgh MoCap \cite{komura2017Edinburgh}&1&Unknown&Unknown\\
        HuMoD \cite{HuMoD}&2&0&No\\
        HumanEva \cite{sigal2010humaneva}&4&0&No\\
        HDM05 \cite{muller2007mocaphdm05} &5&Unknown&No\\
TotalCapture \cite{TotalCaptureTrumbleBMVC2017}&5&Unknown&No\\
        LAFAN1 \cite{harvey2020robustLAFAN1}&5&Unknown&Unknown\\
        GroundLink \cite{han2023groundlink}&7&0&No\\
        SFU Mocap \cite{SFUMocap}&8&0&No\\
        UnderPressure \cite{underpressureMourot22}&10&0&No\\
        Human3.6M \cite{ionescu2013human36m}&11&0&No\\
        PHSPD \cite{phspdzou2020polarization}&12&0&No\\
        ACCAD \cite{ACCAD}&20&0&No\\
        Movi \cite{ghorbani2021movi} &90&0&No\\
        BMLRUB \cite{BMLRUB}&111&0&No\\
        Xia \emph{et al.} \cite{xia2015dataset}&Unknown&Unknown&Yes\\
        BFA \cite{aberman2020unpaired}&1&Unknown&Yes\\
        100style \cite{100style2022}&1&Unknown&Yes\\
        CMU MoCap \cite{CMUMOCAP}&109&Unknown&Yes\\ 
        HUMOTO\cite{lu2025humoto}&1&Unknown&No\\
        PA-HOI\cite{wang2025pa}&2&Unknown&No\\
        \hline
    \end{tabular}
    \caption{Basic information of general-purpose MoCap locomotion datasets.}
    \label{tab:nAllCV}
\end{table}
\begin{table}
    \centering
    \begin{tabular}[\columnwidth]{cc}
    \hline
        Datasets&Old-style forward walking (min)\\
        \hline
        CMU Mocap \cite{CMUMOCAP}&$\sim 2$\\ 
        Xia \emph{et al.}\cite{xia2015dataset}&$\sim 0.75$\\
        BFA \cite{aberman2020unpaired}&$ \sim  4.7$\\
        100style \cite{100style2022}&$ \sim 4.2$\\
        \hline\\
    \end{tabular}
    \caption{Duration of old-style forward walking motions, including a few moments without walking-style arm swings.}
    \label{tab:nOld}
\end{table}
\subsubsection{Motion Variety Within the Same Motor Skills}
The surveyed general-purpose datasets include many different motor skills but have limited motion clips for each, reducing within-skill variety, especially in old-style motions. For example, the CMU MoCap dataset \cite{CMUMOCAP} includes five motion clips related to old-style movements performed by two actors. However, these motion clips (``old man coffee mug/pick up/wait/walk'' and ``elderlyman'') represent different motor skills rather than variations within a single motor skill. The BFA dataset \cite{aberman2020unpaired} contains motion clips of waving hands, but includes few instances of natural arm swings during daily walking. For walking motions, the total walking minutes and the number of participants can still be used as a quantitative measure. However, this measure can be overly optimistic when assessing age-related decline, as only motions paired with normative walking are included to control for performance-related artifacts, assuming that artificial variations across motion styles are comparable.

\section{Fidelity of Old-style Walking Motions}\label{sec:fidelity}
Old-style motions performed by younger actors appear to be a substitute for motions from older adults in several general-purpose datasets. These motions though visually plausible to untrained eyes, can exaggerate part of movement patterns and fail to reflect age-related differences. Age-related differences in clinical community are identified by comparing the same motor skills performed by older and younger adults. To examine whether old-style motions reflect such age-related differences, walking motions are considered. Next, we summarize age-related gait parameters used in clinical community and compared them in old and normative style motion pairs. 

\subsection{Age-related Variables in Walking}\label{subsec:gaitpara}
Gait analysis has been widely adopted by the clinical community to quantify age-related changes. A gait cycle starts at a heel-strike moment and ends at the successive heel-strike moment for the same foot. According to \cite{hollman2011normativephasesrhythm}, kinematic age-related variables from \cite{hausdorff2005gaitvariability,BOYER201763gaitmeta,kneeromdifference,Begg01012006ankleanglePhase,abbass2013gaitkinematicoverview,aginglessefficient} are identified in five domains: rhythm, phase, pace, balance control, and variability.

Rhythm variables consist of temporal variables such as stride time, step time, and cadence. \textit{Stride time} is the duration of a gait cycle. Time spent on a single step is defined as \textit{step time}. \textit{Cadence} is the number of steps per minute. In this work, the stepcount algorithm is based on periodic changes in foot height and visual observation. 

Phase variables characterize the temporophasic divisions of the gait cycle. A gait cycle can be divided to stance and swing phases. The two phases can be identified using heel- strike and toe-off moments, whereas single and double limb support phases require more granular information on foot-floor interaction. The \textit{duration of each phases normalized to stride time} can be used to assess gait cycle dynamics.

\begin{figure}[h]
\centering
\includegraphics[width=0.7\columnwidth]{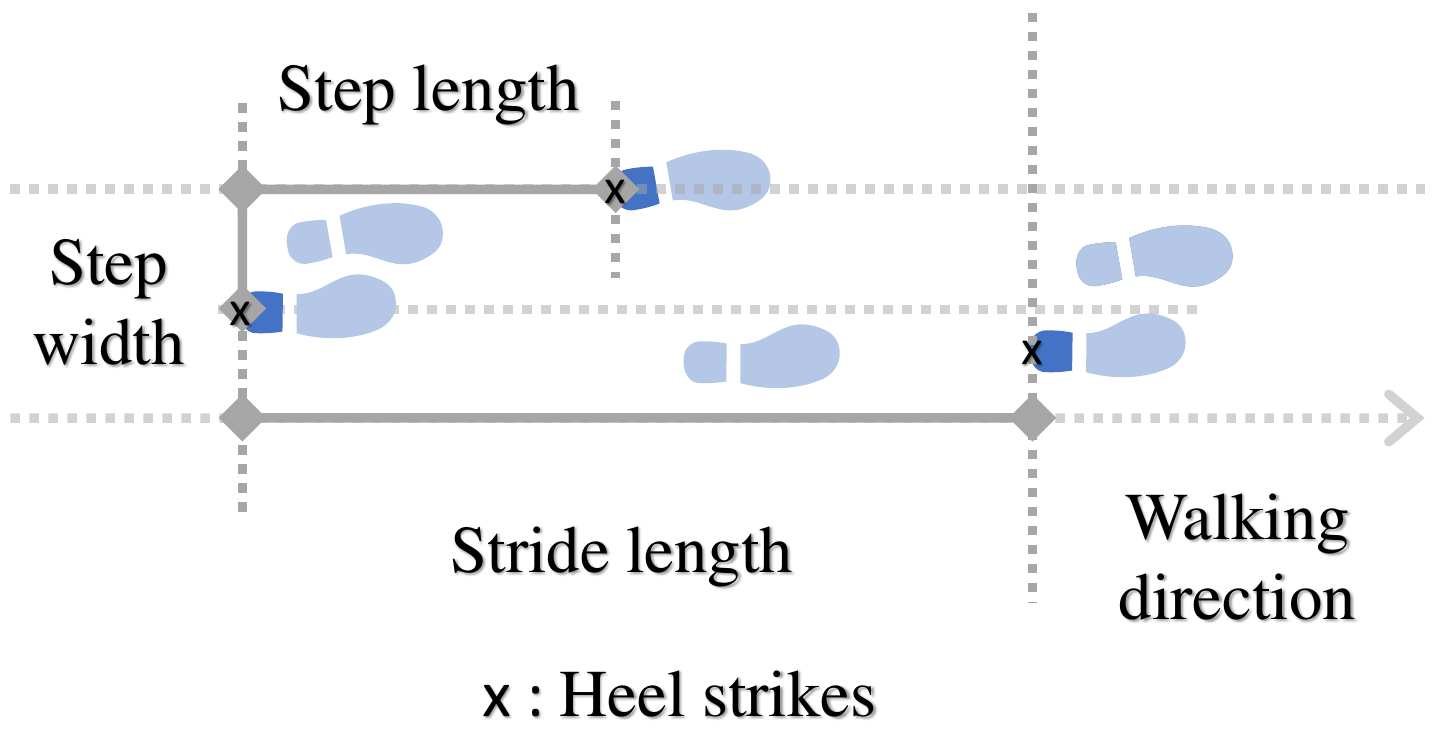}
\caption{Basic gait parameters in a gait cycle.}
\label{fig:gaitvat}
\end{figure}

Pace variables includes \textit{step length}, \textit{stride length}, and \textit{gait speed}. The distance between two successive heel prints along the walking direction is the step length. The distance between the same foot at heel-strike moments in a gait cycle is called the stride length. Gait speed can be measured by the distance traveled along the walking direction divided by the time taken. Knowing foot positions at heel-strike moments allows the calculation of basic gait parameters in a gait cycle, as illustrated in Fig.~\ref{fig:gaitvat}.

Balance control variables such as \textit{step width}, the \textit{margin of stability} \cite{yamaguchi2022effectsdynamicmeasures}, and \textit{joint angles} are related to balance control during walking. \textit{Step width} is defined as the mediolateral distance between successive foot placements, measured orthogonal to the walking direction. The calculation of the \textit{margin of stability} in walking requires identifying the contact points on foot or footwear with the ground as well as estimating the location of the center of mass \cite{yamaguchi2022effectsdynamicmeasures}. Joint angles at heel-strike and toe-off moments are informative. 

Variability is defined as the \textit{standard deviations} of aforementioned variables such as step width and stride length. Moreover, the \textit{range of motion} (RoM) of joints such as thorax and knee joint are considered. RoM is the difference between the maximum and minimum angular displacements of a joint during a movement cycle. 

Generally, aging causes negative changes in gait parameters in all five domains. However, confounding factors such as walking intentions, footwear, and surface conditions, can result in changes as well independent of age. For example, trunk kinematics vary when participants were instructed to swing their arms in different patterns \cite{2020armswingRole}. Gait speed affects stride length, cadence, and joint kinematics. Instructing someone to walk faster or slower can induce compensatory changes. To accurately isolate age-related differences, it is essential to control for confounding factors, whereas accounting for potential noise and data scarcity. For instance, when individuals walk at self-selected speeds, reductions in gait speed and knee RoM are frequently observed as indicators of aging, provided that these reductions are not the consequence of noise.

\subsection{Age-related Gait Parameters in Old-style Motions}\label{subsec:limitation}
To select suitable gait parameters to evaluate whether old-style walking motion reflects age-related differences, we consider the following criteria:
\begin{itemize}
    \item Independence from the walking protocols. A walking protocol is supposed to measure quantities of interests while controlling other confounding factors. Controlled gait parameters should be excluded from the consideration of age-related parameters, e.g., gait speed when participants were instructed to walk at a predefined speed.
    \item Robustness to data scarcity. The selected parameters should remain discriminative under limited data. For example, stride variables cannot be reliably estimated if only few gait cycles are available. In contrast, step variables can be better estimated even a few gait cycles by taking advantage of two steps per stride. The symmetry between left and right legs is not age-related during straight walking, thus no extra error is introduced when analyzing, e.g., three consecutive steps \cite{symmetrynotfeaturePATTERSON2012590}.
    \item Accuracy in capturing age-related differences. The selected parameters are expected to remain discriminative under noisy data conditions and modeling approximations. When foot is modeled as a single point rather than including heel and toes, the estimation of heel-strike and toe-off moments becomes inaccurate.  
\end{itemize}

Next, we apply these criteria to examine the suitability of age-related parameters discussed in Section \ref{subsec:gaitpara}. 
Rhythm variables are robust to noise and can be generalized across various walking protocols such as walking straight or following a figure-of-8 test (F8WT) \cite{abbass2013gaitkinematicoverview,zancan2021basicCMUfigureof8}, but they can be sensitive to data scarcity. For instance, a new step may not be counted even if it is $90\%$ complete when estimating cadence. This error can be particularly significant in segments that only have a few steps.
The accuracy of phase variables is sensitive to the accuracy of heel-strike and toe-off moments. Pace variables are independent from walking protocols when participants walking along the same trajectories. Gait speed remains discriminative in daily walking scenarios, allowing variations in walking trajectories. Among the balance control variables, the accuracy of step width depends on the detection of heel-strike moments. Variables such as center of mass requires information other than MoCap recordings. Among gait parameters characterizing variability, step width is sensitive to the variations in walking trajectories. Knee RoM is robust to noise. Conventional MoCap errors in estimating hip joint flexion and abduction angles can reach 7.2 and 4.3 degrees during gait \cite{lopez2024hipaccuracy}, respectively. The general source of error obscures age-related differences in hip angles.

In summary, suitable age-related variables are gait speed and knee RoM. The feasibility of rhythm variables depends on the availability of sufficient data. Pace variables and step width are capable to show age-related differences if the variations of walking trajectories can be controlled and the detected heel-strike moments represent a consistent phase in gait cycles. 

\subsection{Assessing Old-style Motions in Existing MoCap Datasets}
This section evaluates the fidelity of the old-style motions in four MoCap datasets \cite{xia2015dataset,CMUMOCAP,100style2022,aberman2020unpaired}.

\begin{figure*}[h!]
    \centering
    \begin{subfigure}[]{0.24\textwidth}        \includegraphics[width=\linewidth]{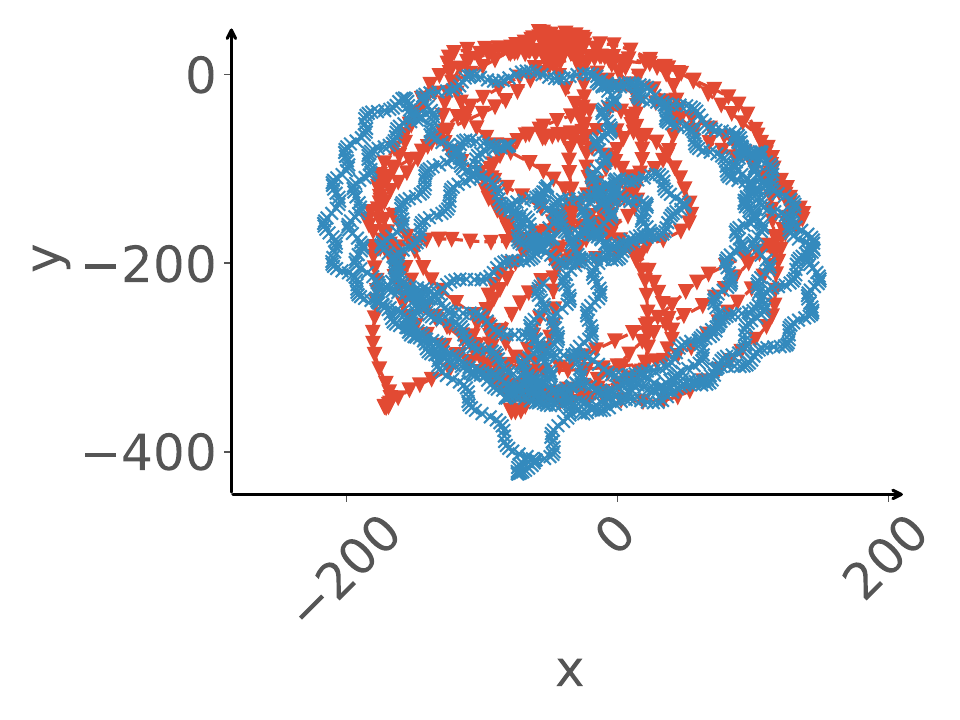}
        \caption{100style, daily walking}
        \label{subfigstyle100traj}
    \end{subfigure}\hfill
    \begin{subfigure}[]{0.24\textwidth}
\includegraphics[width=\linewidth]{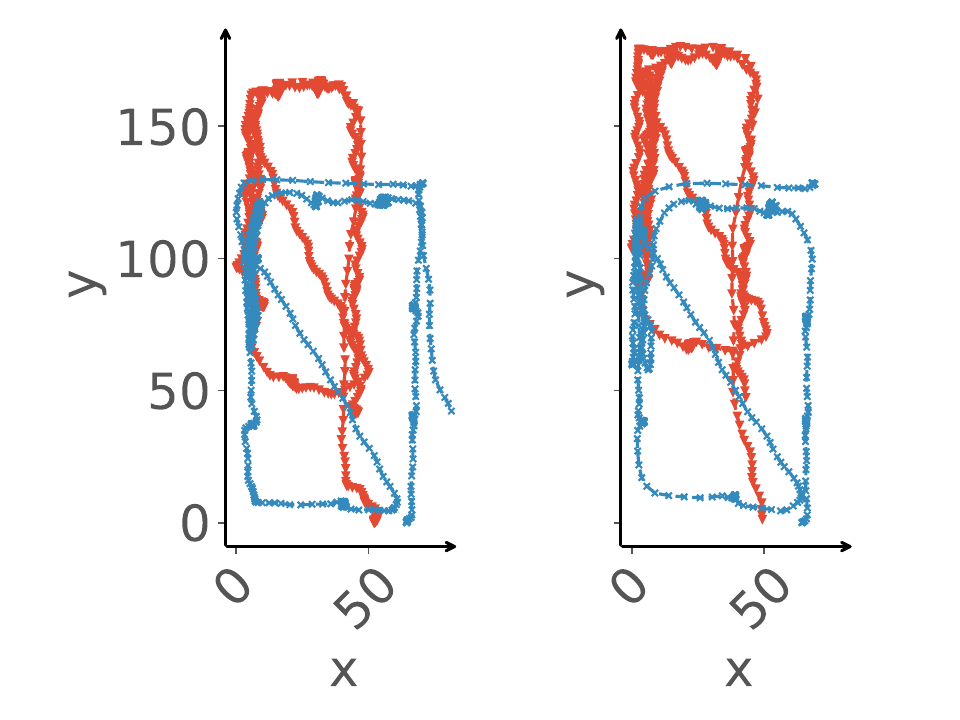}
        \caption{BFA, preferred speed}
        \label{subfigbfatraj}
    \end{subfigure}\hfill
    \begin{subfigure}[]{0.24\textwidth}
\includegraphics[width=\linewidth]{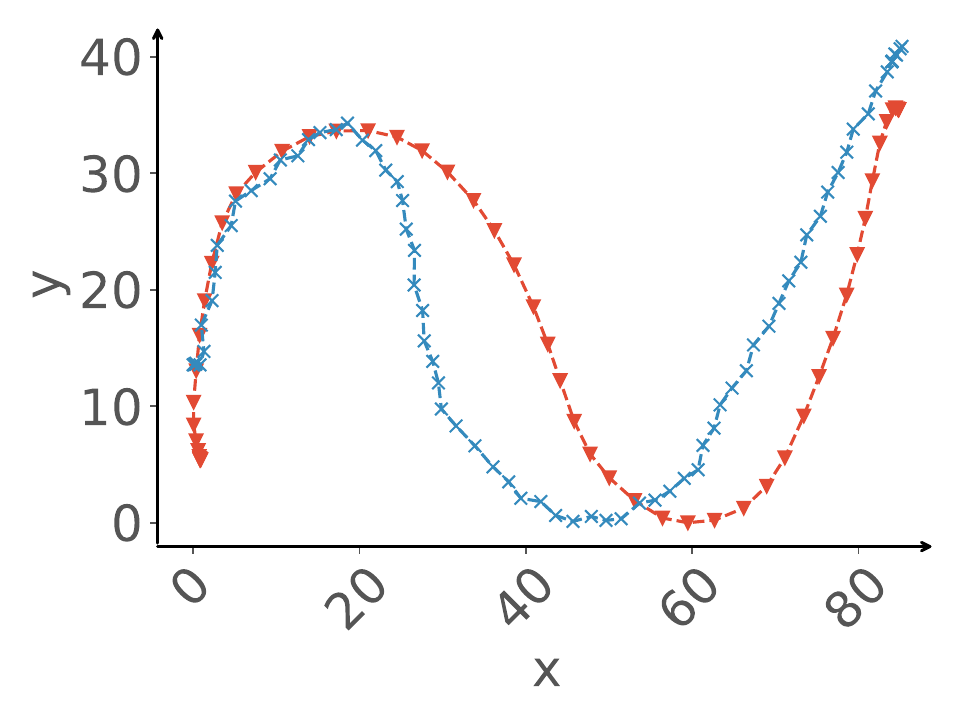}
        \caption{CMU MoCap, F8WT}
        \label{subfigcmutraj}
    \end{subfigure}\hfill
       \begin{subfigure}[]{0.24\textwidth} \includegraphics[width=\linewidth]{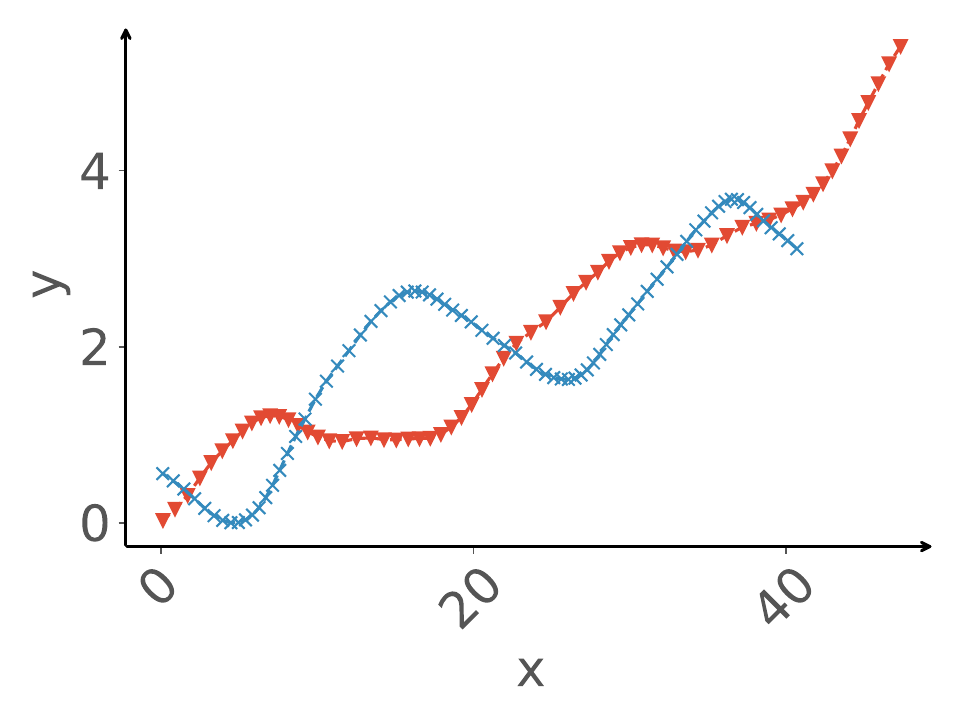}
        \caption{Xia \emph{et al.}, preferred speed}
        \label{subfigxiatraj}
    \end{subfigure}\hfill
   
    \caption{Walking trajectories of old-style forward walking motions and their control groups. Old-style motion and normative walking are plotted in blue and red, respectively. The units of the x and y axes are equal. The walking protocols included are daily walking, walking at a preferred speed, and F8WT walking.}
    \label{fig:traj}
\end{figure*}

\subsubsection{Determining Gait Parameters}
Among all the gait parameters used in clinical community, some are excluded if they are of low accuracy or sensitive to data scarcity and variations in walking protocols, such as differences in walking trajectories. We selected segments from walking trials that contained only steady-state walking without substantial upper limb movements. Additionally, upon visual inspection of the walking trajectories in Fig.~\ref{fig:traj}, segments exhibiting sharp turns were excluded. The walking protocols used to select suitable gait parameters are daily walking, walking at a preferred speed, and F8WT walking.

\begin{figure}[h!]
    \centering
    \begin{subfigure}[b]{0.25\textwidth}
        \centering
        \includegraphics[width=\textwidth]{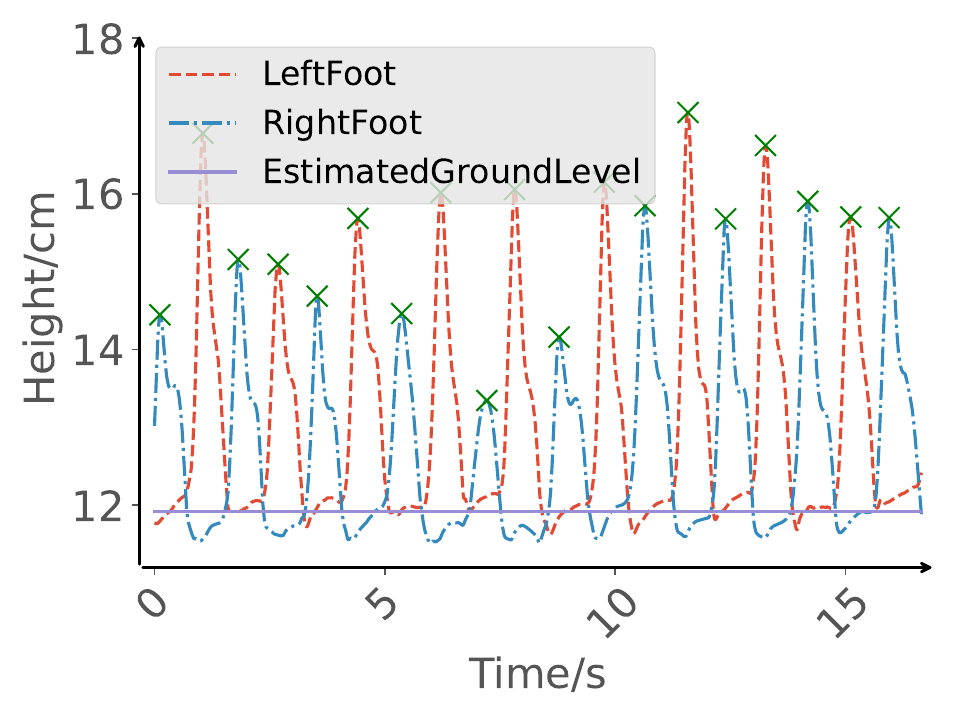}
        \caption{100style}
        \label{subfig:style100}
    \end{subfigure}
    \hfill
    \begin{subfigure}[b]{0.25\textwidth}
        \centering
        \includegraphics[width=\textwidth]{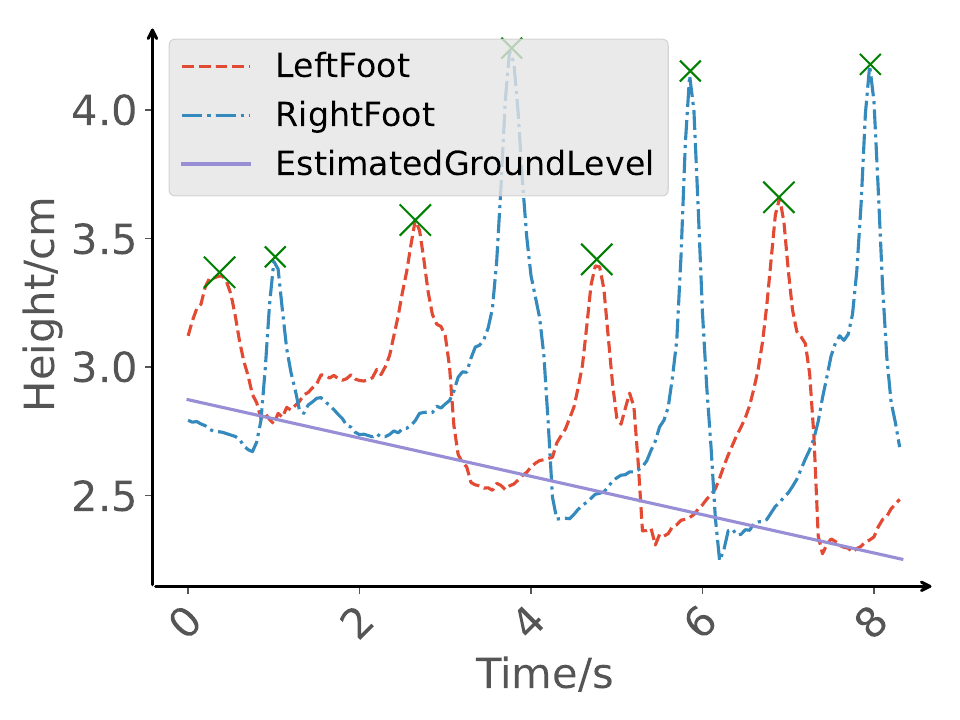}
        \caption{BFA}
        \label{subfig:bfa}
    \end{subfigure}
    \caption{Feet height and step counting in old-style forward walking records. The x- and y-axis represent time and height values, respectively. The green crosses are used to count steps.}
    \label{fig:bfastepcount}
\end{figure}

Estimating heel-strike events and identifying gait cycles are essential steps in calculating most of the aforementioned gait parameters. The accuracy of these processes varies across datasets. It was verified through our visual inspection. As shown in Fig. ~\ref{fig:bfastepcount}, the number of gait cycles can be identified ideally using peak detection methods. The peak values occur at consistent phases across gait cycles. The feet-floor interaction, defined by the height of feet and ground level, cannot be accurately estimated if foot is represented by a single joint in motion data. Annotating heel-strike moments with visual inspection is subjective and may affect gait parameters. 

The heel-strike moments for walking motions in Xia \emph{et al. }\cite{xia2015dataset} were annotated based on a consistent phase of the gait cycle and verified through visual inspection. The lowest feet heights in each gait cycle keep changing during the forward walking motion clips from both 100style \cite{100style2022} and BFA datasets \cite{aberman2020unpaired}. The estimated ground level from the first gait cycle is severely inaccurate for detecting heel-strike moments in subsequent gait cycles, and even for a consistent phase within the same cycle. Thus, gait parameters requiring heel-strike moments were excluded when evaluating 100style \cite{100style2022} and BFA datasets.

The walking trajectories between normative and old-style walking motions are different in 100style \cite{100style2022} and the CMU MoCap dataset \cite{CMUMOCAP}. To minimize the influence of such variations, the selected gait parameters for age-related analysis are gait speed, cadence, and knee RoM. The inclusion and exclusion of age-related variables are summarized in TABLE~\ref{tab:gaitselection}. 

\begin{table*}[h]
    \centering
    \begin{tabular}{cccccc}
    \hline
        Variables&\multicolumn{4}{c}{Inclusion/Exclusion (\checkmark / \ding{55})}&Expected Changes In old-style Motions\\
       &Xia \emph{et al.} \cite{xia2015dataset}&CMU Mocap \cite{CMUMOCAP}&100style \cite{100style2022}&BFA \cite{aberman2020unpaired}& $/$Reasons for Exclusion\\
       \hline
        Gait speed, mean&\checkmark &\checkmark &\checkmark &\checkmark & smaller\\ 
        Cadence (steps per min)&\checkmark &\checkmark &\ding{55}&\ding{55}&smaller/protocol dependence\\
        Avg step width&\checkmark&\ding{55}&\ding{55}&\ding{55}&larger/protocol dependence\\
        Avg step length&\checkmark&\ding{55}&\ding{55}&\ding{55}& smaller/protocol dependence\\
        Gait speed, std&\checkmark &\checkmark &\checkmark &\checkmark& larger\\
        Stride time variability&\ding{55}&\ding{55}&\ding{55}&\ding{55}&data scarcity\\
        Step length variability&\checkmark&\ding{55}&\ding{55}&\ding{55}& larger/inadequate accuracy\\
        Step width variability&\checkmark &\ding{55}&\ding{55}&\ding{55}& larger/protocol dependence\\
        Knee RoM&\checkmark &\checkmark &\checkmark &\checkmark& smaller\\
        Ankle RoM&\ding{55}&\ding{55}&\ding{55}&\ding{55}& Inadequate accuracy\\
        Hip RoM&\ding{55}&\ding{55}&\ding{55}&\ding{55}& Inadequate accuracy\\
        Ankle angles at heel-strike/toe-off moments&\ding{55}&\ding{55}&\ding{55}&\ding{55}&Inadequate accuracy\\
        Dynamic balance measures \cite{yamaguchi2022effectsdynamicmeasures}& \ding{55}&\ding{55}&\ding{55}&\ding{55}& Inadequate accuracy\\
         \hline
    \end{tabular}
    \caption{Inclusion/Exclusion of variables for gait analysis and the expected age-related differences of old-style and normative walking motion pairs in each datasets.}
    \label{tab:gaitselection}
\end{table*}

\subsubsection{Results} As shown in TABLE~\ref{tab:gaitalldata}, old-style motions exhibit lower gait speeds and smaller RoM values of knee flexion compared to younger motions in the datasets. These differences are consistent with clinical findings.

However, several key gait parameters in old-style motions diverge from age-related gait patterns. In Xia \emph{et al.}'s dataset, for instance, the step width in older-style motions is approximately half that of the younger group, whereas the stride lengths remain similar. However, among real older adults, it is more common to have a decreased gait speed without having a significant difference in step width \cite{elble1991stridelowergaitbysteplength,ostrosky1994comparisonlowergaitbysteplength}. Narrow steps require greater balance control, whereas a wider stance is a strategy used to enhance stability. Similarly, datasets such as CMU MoCap, BFA, and 100style report substantially lower variability in gait speed for older-style motions, around $30\%$ that of the younger group. This reduced variability implies better balance control, which directly contradicts real-world observations. Furthermore, the cadence of older-style motions is reported to be higher than that of younger individuals in CMU, again deviating from expected age-related declines.

\begin{table*}[h!]
    \centering
    \begin{tabular}{cccccccc}
    \hline Dataset&Style&Gait speed, mean$\pm$std&Step length, mean$\pm$std&Step width, mean$\pm$std&KneeROM (deg)&Cadence\\
         \hline
        \multirow{2}{*}{Xia \emph{et al.} \cite{xia2015dataset}}&Old&0.167$\pm$0.033&12.34$\pm$0.5783&\textbf{1.168$\downarrow$}$\pm$0.294&26.10&94.35\\
        &Normative&0.200$\pm$0.024&12.88$\pm$0.445&2.039$\pm$0.217& 38.47&105.71\\
                 \hline
         \multirow{2}{*}{CMU MoCap \cite{CMUMOCAP}}
         &Old&0.055$\pm$\textbf{0.009$\downarrow$}&\ding{55}&\ding{55}&14.59&\textbf{144.18$\uparrow$}\\
         &Normative&0.173$\pm$0.024&\ding{55}&\ding{55}&42.47&109.25\\
         \hline
         \multirow{2}{*}{BFA \cite{aberman2020unpaired}}
         &Old&0.061\textbf{$\pm$0.015$\downarrow$}&\ding{55}&\ding{55}& 21.30&\ding{55}\\
         &Normative&0.090$\pm$0.037&\ding{55}&\ding{55}& 36.76&\ding{55}\\
         \hline
         \multirow{2}{*}{100style \cite{100style2022}}&Old&0.399\textbf{$\pm$0.120$\downarrow$}&\ding{55}&\ding{55}&33.29&\ding{55}\\
         &Normative&1.217$\pm$0.381&\ding{55}&\ding{55}&41.02&\ding{55}\\
         \hline
    \end{tabular}
    \caption{Evaluation of age-related differences in old-style and normative motions within each dataset. The units of gait speed, step length, and step width depend on the spatial unit used in the motion records. Variables that are not comparable or show inconsistent age-related differences of old-style motions are highlighted in bold.}\label{tab:gaitalldata}
\end{table*}

\begin{figure}[h]
    \centering
    \begin{subfigure}[b]{0.25\textwidth}
        \centering
        \includegraphics[width=\textwidth]{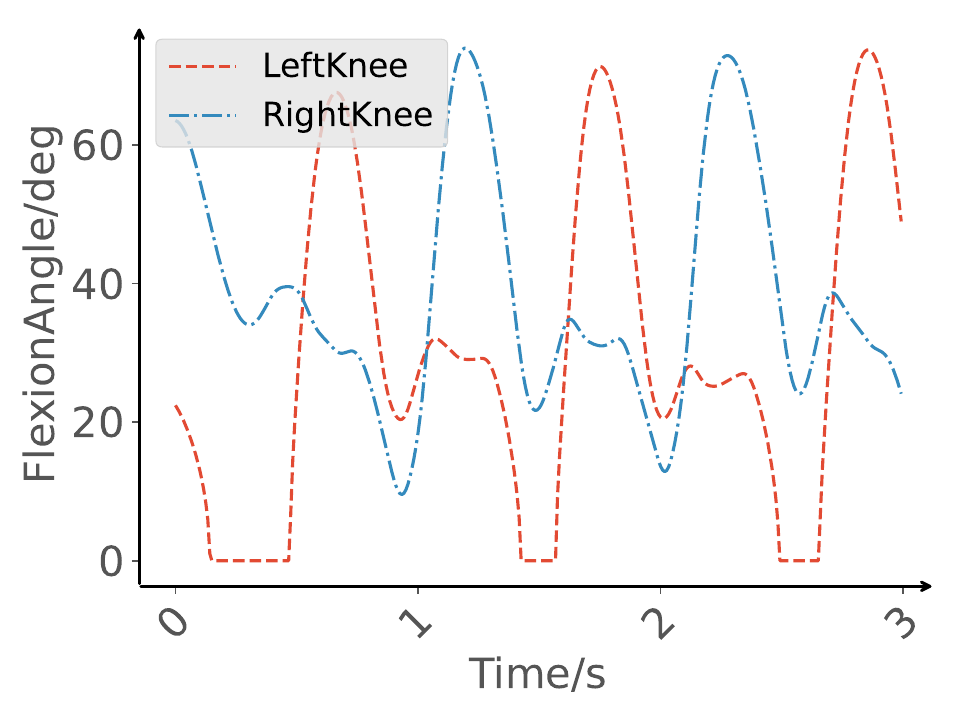}
  \caption{Normative walking, take 29}
  \label{fig:sub1}
    \end{subfigure}
    \hfill
    \begin{subfigure}[b]{0.25\textwidth}
        \centering
        \includegraphics[width=\textwidth]{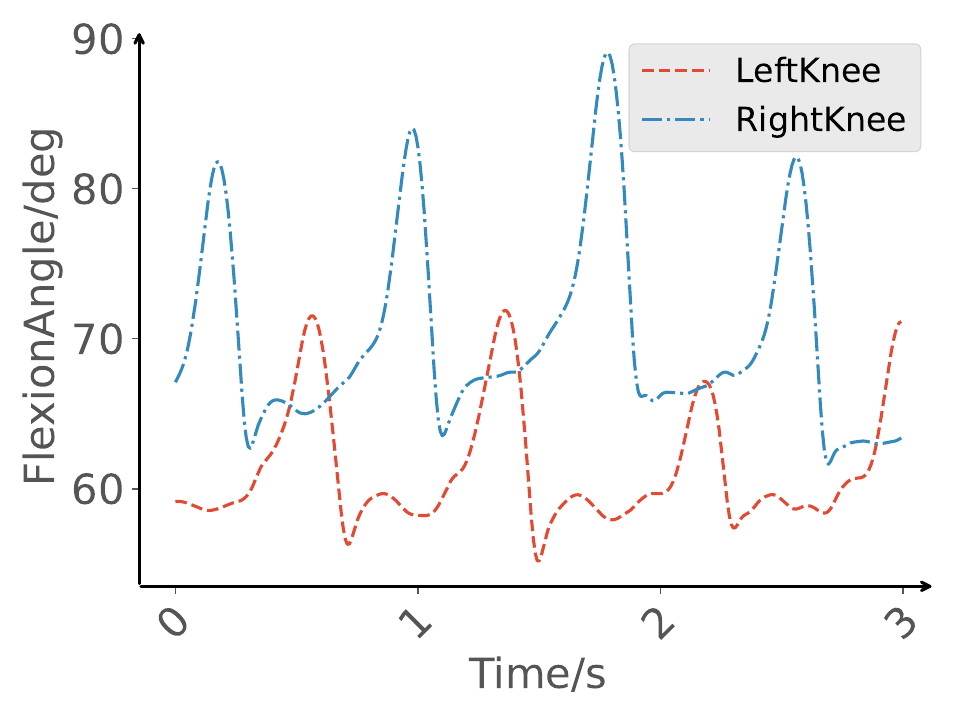}
  \caption{Old-style walking, take 33}
  \label{fig:sub2}
    \end{subfigure}
\caption{Visualization of smoothness using joint kinematics from walking motions from subject 137 of the CMU dataset.}
\label{fig:test}
\end{figure}

Visualizing the gait parameters also make it intuitive to assess the quality of old-style motions. For example, Fig.~\ref{fig:test} compares the knee flexion kinematics in normative and old-style walking clips from the CMU dataset. During the short segments analyzed, the walking trajectories are approximately straight. The asymmetry in knee flexion angles is found. Moreover, the maximum flexion angles during gait cycles in old-style walking are approximately 20 degrees greater than those observed in normative walking motions. Visualizing the gait parameters makes the non–age-related differences evident, even though the walking motions are too short to reveal significant differences.
\subsubsection{Discussion} The limitation of the fidelity assessment is data scarcity, restricting our method to walking motions in a few datasets. We had to confine comparison to motion pairs since the unnaturalness from acting is difficult to quantify. The results cannot be generalized across datasets and may be weakened by differences among walking protocols used in these datasets and those used in clinical settings. 

For dataset creators who intend to use younger actors as substitutes for older adults, we recommend determining age-related parameters of older adults to replicate beyond mere visual similarity, putting more sensors on body parts to guarantee the accuracy of these parameters, and constructing comparable motion pairs for fidelity assessment. 

\section{Conclusion}\label{sec:conclusion}
This work examined existing MoCap locomotion datasets and investigated their demographics, motion variety, and fidelity. Our survey included 41 MoCap locomotion datasets and shows that limited data collected from older adults are publicly available. Clinical datasets are characterized by limited motor skills and motion variety, a scarcity of full-body motion recordings, and a predominance of in-lab walking tasks. General-purpose datasets have no documented older adults, with a few containing old-style motions. 

The fidelity of available old-style motions has been found questionable according to age-related gait parameters. Old-style walking motions demonstrated lower speed variability and higher cadence compared to the actors’ normative walking. This strongly suggests a mismatch between the intended portrayal of age-related gait and the performance. Moreover, visualization of the age-related variables reveals the unnaturalness of acting when motions have few gait cycles. Future work involves extending the pipeline to other motor skills and incorporating additional sensor types beyond MoCap. This will include developing age-related variables that are robust to data scarcity and non-age-related variations within motor skills.

\section*{Acknowledgment}
This work is supported partly by NSERC sMAP CREATE and NSERC Discovery grants.

\bibliographystyle{IEEEtran}
\bibliography{IEEEexample}

\end{document}